\documentclass[twocolumn,showpacs,preprintnumbers,amsmath,amssymb]{revtex4}

\def\be{\begin{equation}}
\def\ee{\end{equation}}

\def\be{\begin{equation}}
\def\ee{\end{equation}}
\def\beq{\begin{equation}}
\def\eeq{\end{equation}}

\def\be{\begin{equation}}
\def\ee{\end{equation}}

\def\mpl{M_{Pl}}

\def\calM{{\cal M}}
\def\calB{{\cal B}}

\usepackage[dvipdfmx]{graphicx}
\usepackage{here}
\usepackage{comment,braket}
\usepackage{epsf}
\usepackage{amsmath}
\usepackage{graphics}
\usepackage{amsfonts}
\usepackage{amssymb}
\usepackage{latexsym}
\usepackage{color}
\input{colordvi.tex}
\usepackage{feynmp}

\begin{document}
\preprint{RESCEU-1/16}

\title{Creation of an inflationary universe out of a black hole}

\author{Naritaka Oshita$^{1,2}$}
\author{Jun'ichi Yokoyama$^{1,2,3}$}
\affiliation{
  $^1$Research Center for the Early Universe (RESCEU), Graduate School
  of Science,\\ The University of Tokyo, Tokyo 113-0033, Japan
}
\affiliation{
  $^2$Department of Physics, Graduate School of Science,\\ The University of Tokyo, Tokyo 113-0033, Japan
}
\affiliation{
  $^3$Kavli Institute for the Physics and Mathematics 
of the Universe (Kavli IPMU), WPI, UTIAS,\\
The University of Tokyo, Kashiwa, Chiba 277-8568, Japan
}

\date{\today}

\begin{abstract}
We discuss a two-step mechanism to create a new inflationary domain
beyond a wormhole throat which is created by a phase transition around
an evaporating black hole.  The first step is creation of a false vacuum
 bubble with a thin-wall boundary by the thermal effects of Hawking
 radiation.  
Then this wall induces a quantum tunneling to create a
 wormhole-like configuration.  
As the space beyond the wormhole throat can expand exponentially, 
being filled with false vacuum energy,
this may be interpreted as creation of another inflationary universe in the
final stage of the black hole evaporation.
\end{abstract}

\pacs{98.80.Bp,04.62.+v,98.80.Qc,04.70.-s}

\maketitle


Inflation in the early universe provides answers to a number of
fundamental questions in cosmology such as why our Universe is big, old,
full of structures, and devoid of  unwanted relics
predicted
by particle physics models \cite{Sato:2015dga}.  Furthermore, despite the great
advancements in precision observations of cosmic
microwave background (CMB) radiation, there is no  observational
result that is in contradiction with inflationary cosmology so far \cite{Akrami:2018odb,Hinshaw:2012aka}.

Inflationary cosmology has also revolutionized our view of the
cosmos, namely, our Universe may not be the one and the only entity but
there may be many universes.  Indeed already in the context of the old 
inflation model \cite{Sato:1980yn,Guth:1980zm},
 Sato and his collaborators found possible production of child (and grand
child...) universes \cite{KSato:1981ptp,KMaeda:1982plb,KSato:1982plb}. 

Furthermore, if the observed dark energy consists of a cosmological
constant $\Lambda$, our Universe will asymptotically approach the de
Sitter space which may up-tunnel to another de Sitter universe with a
larger vacuum energy density 
\cite{Hawking:1981fz,Lee:1987qc,Phys.Rev.D76.103510,Phys.Rev.D.57.2230,Oshita:2016oqn}
to induce inflation again to repeat the
entire evolution of another inflationary universe.  
In such a recycling universe scenario, the
Universe we live in may not be of first generation, and we may not need
 the real beginning of the cosmos from the initial
singularity \cite{Phys.Rev.D.57.2230}.

In this context, so far only a phase transition between two pure de Sitter
space has been considered \cite{footnote1}.  However, phase transitions which we
encounter in daily life or laboratories are usually induced around some
impurities which act as catalysts or boiling stones.  In cosmological
phase transitions, black holes may play such roles.  
In this manuscript we discuss a cosmological phase transition around an evaporating
black hole to show that a wormhole-like configuration with an inflationary domain beyond the throat 
may be created after the transition. 

The study of a phase transition around a black hole was pioneered by Hiscock \cite{Hiscock:1987}.  
More recently, 
Gregory, Moss and Withers
revisited the problem \cite{Gregory:2014JHEP}.  
They have observed that the black hole mass may
change in the phase transition and calculated the Euclidean action
taking conical deficits into 
account \cite{Gregory:2014JHEP,Phys.Rev.Lett.115.071303,JHEP.1508.114}.
Moreover, a symmetry restoration activated by 
Hawking radiation \cite{Hawking:1974rv,Hawking:1974sw} near a microscopic
black hole has been investigated by Moss \cite{Moss:1984zf}.


We consider a high energy field theory of a scalar field $\phi$ whose
potential allows a thin-wall bubble solution of
 a metastable local minimum at $\phi=0$ with the energy
density $\epsilon^4$ surrounded by the true vacuum
with a field value $\phi_0$  where the mass square is given by
$m^2$.  In such a theory Moss \cite{Moss:1984zf}
argues that the symmetry is restored in the vicinity of the black hole
horizon inside a thin wall bubble
as the Hawking temperature, $T_H=\mpl^2/(8\pi M_+)$, reaches
the mass scale of the theory.
Here $M_+$ and $\mpl$ are the black
 hole mass and the Planck mass, respectively.
In the presence of
 plausible couplings of the relevant fields, he shows that 
 the medium inside the bubble, where fields coupled to $\phi$ are
massless, is thermalized with a temperature $T $ which is substantially
smaller than $m\sim T_H$.  Then the free energy of the bubble configuration
 is given by
\begin{equation}
  F(r,T)=\frac{4}{3}\pi r^3\epsilon^4+4\pi r^2\sigma -\frac{\pi}{18} q
  {\tilde m}^2T^2r^3
  \label{freeenergy}
\end{equation}
as a function of its radius $r$ and $T$, 
where $\sigma$ is the surface tension of the wall and
$\tilde{m}^2$ denotes sum of the mass squared of species which receive
a mass from $\phi$ outside the bubble.  For simplicity we assume
${\tilde m}$ is of the same order of $m$ and omit the tilde hereafter.
Here $q$ is related to the scattering
 parameter $C$ defined by Moss  \cite{Moss:1984zf} as
 $q\equiv (192\pi^2 C)^{-2/3}$, which can take a value of order of unity
 or even larger.

The relation between the thermalized temperature $T$ and the bubble
radius $r_w$ is obtained by  solving the Boltzmann equations for the
radiated beam particles and thermalized medium with the boundary
condition that only particles with energy larger than $m$ would escape
the bubble wall, which reads
\begin{equation}
 \frac{1}{216}q^{-3/2}T^3r^3 + 48mTr^2e^{-\beta m}=1,
  \label{boundary}
\end{equation}
at $r=r_w$ with $\beta\equiv T^{-1}$.

The radius of the wall $r_w$ is obtained by minimizing 
the free energy (\ref{freeenergy}) under the condition (\ref{boundary}).  For example, 
when the inequality 
\begin{equation}
mr_w \gg 10^4q^{2/3}(\beta m)^2e^{-\beta m}  \label{inequality}
\end{equation}
is satisfied and the first term dominates the left hand side of (\ref{boundary}),
we find $T=6\sqrt{q}/r_w$, so that the free energy is minimized at
\begin{equation}
r=r_w = 
\sqrt{\frac{3q}{2}}\frac{m}{\epsilon^2}.
\end{equation}

For consistency of this solution with (\ref{inequality}),
$\epsilon$ and $m$ must satisfy
\begin{equation}
 \frac{m}{T}=\frac{1}{6\sqrt{2}}\frac{m^2}{\epsilon^2} \gtrsim 10,  \label{mT}
\end{equation}
which we assume hereafter.  Then the thin wall condition
$mr_w = {\displaystyle \sqrt{\frac{q}{2}}\frac{m^2}{\epsilon^2}} \gg 1$ is
naturally satisfied.

Under the condition (\ref{mT}) thermal energy inside the bubble is subdominant
compared with $\epsilon^4$, so
the geometry inside the bubble can be described by the Schwarzschild de
Sitter metric.  Furthermore,
as the radiation temperature increases in association with the increase of the
Hawking temperature, more high energy particles, which
escape from the bubble and do not contribute to support the wall, are
created to lower the effect of the radiation pressure.  Thus, contrary to naive expectation,
thermal effects on the created bubble become less important as the  temperature 
increases, which can be also understood from the inequality $dr_w/dT <0$ 
 derived from
(\ref{boundary}).
 
  Thus the system
can be approximated by  a spherically
symmetric thin wall with tension $\sigma$ 
separating outside Schwarzschild geometry with mass
parameter $M_+$ and inside
Schwarzschild de Sitter geometry with vacuum energy density 
$\epsilon^4\equiv 3\mpl^2 H^2/(8\pi)$
whose mass parameter we denote by $M_-$.

We use the  equation of motion of the wall 
 obtained by Israel's junction condition
to discuss quantum tunneling of
the bubble to show that the final state is a wormhole-like configuration.
Beyond the throat
is a false vacuum state which inflates to create another big universe.
Then one may regard that the final fate of an evaporating black hole is
actually another universe.
We do not take thermal effects to tunneling into account, as they would 
only enhance the tunneling rate.

\if
In the present manuscript we study the stability of the false vacuum 
nucleated around the
microscopic black hole assuming that the high energy theory of elementary
interactions accommodate a false vacuum with energy density 
$U= M_X^4 \equiv 3\mpl^2 H^2$, where $\mpl$ is the reduced 
Planck scale, and that transition between this state and the
current vacuum state may take place through a thin wall bubble nucleation
with its surface tension $\sigma$ when the black hole mass decreases to $\sim M_{Pl}^2/M_X$
due to the Hawking radiation \cite{Hawking:1974rv,Hawking:1974sw}
(See Fig.\ \ref{112001}). 
We assume the energy scale $M_X$ is somewhat smaller than 
the typical
grand unification scale $M_{\rm GUT}\sim 10^{16}$GeV
on the basis of the constraints imposed on the energy scale of inflation from 
B-mode polarization of CMB \cite{Ade:2015tva}.
As a result we show that once the false vacuum bubble is nucleated
around the black hole, it may immediately and spontaneously
decay to a more stable structure with a worm hole-like throat.
\fi

\if
We study how such a configuration may be created from the thin wall false vacuum bubble
nucleated around the microscopic black hole by calculating
the tunneling rate of the false vacuum bubble in the Euclidean picture.
Since the energy
scale of the false vacuum is presumably much larger than that of the
current dark energy, we neglect the latter.  To be more specific, we
consider the case a false vacuum bubble with its radius $R$
is nucleated around a Schwarzschild black hole with mass $M_-$.
\fi

\begin{figure}[b]
  \begin{center}
    \includegraphics[keepaspectratio=true,height=55mm]{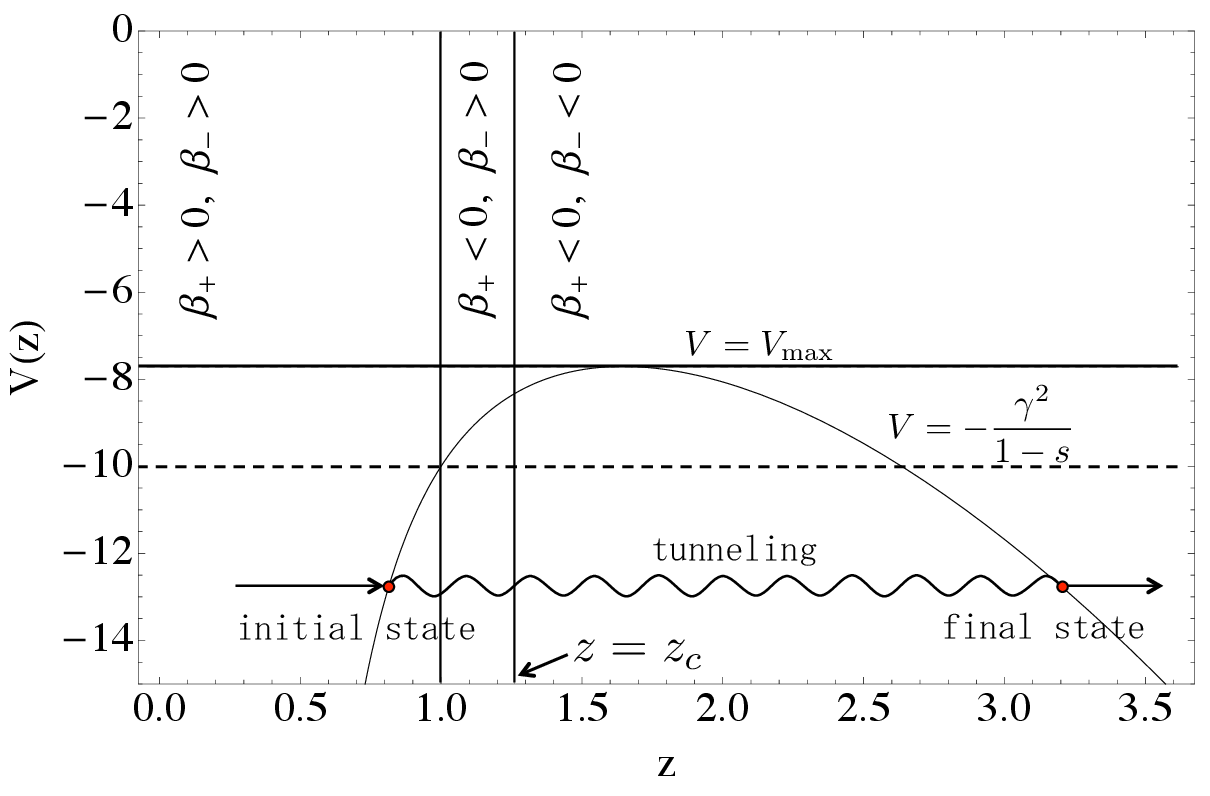}
  \end{center}
  \caption{
Shape of the potential $V(z)$ as a function of $z$ with  $s=0.9$.
We have taken $\gamma=1$ for illustrative purpose, although we actually
expect $\gamma \ll 1$ for $\phi_0 \ll M_{\rm Pl}$. 
$\beta_+$ changes its sign at $z=1$, 
and $\beta_-$ at $z=(1-\gamma^2/2)^{-1/2}\equiv z_c$.
\if
Physical separation between $z=z_c$ and $z=1$ is given by $\gamma^2 r_w/4$
which is smaller than the width of the wall, $\sim 1/m$, for realistic values of $\gamma$.
Hence one can regard that $\beta_\pm$ changes their signs practically  at the same radius 
under the thin wall approximation.
When $E$ takes the value lower than $-\gamma^2/(1-s)$ (dashed line),
the bubble wall is located in the region where  $\beta_{\pm}$ both take
a positive sign before the tunneling commences, 
whereas they both become negative after the tunneling, which implies the
transition from a black hole to a wormhole.
\fi
}%
  \label{112103}
\end{figure}

We label the inner Schwarzschild de Sitter  geometry with a suffix
$-$ and outer
Schwarzschild geometry with a suffix $+$. 
Then the outer and inner metrices are given  by
\begin{eqnarray}
&&ds^2 = -f_{\pm}(r) dt^2 +
 \frac{ dr^2}{f_{\pm}(r)}
 + r^2 d\Omega^2,\\
&&f_+(r) \equiv 1- \frac{2GM_+}{r}, \ f_-(r) \equiv 1-\frac{2GM_-}{r} -
 H^2 r^2.
\nonumber
\end{eqnarray} 
We describe the wall trajectory in terms of the local coordinates
$(t_{\pm}(\tau),r_{\pm}(\tau),\theta,\varphi)$ on each side depending on
 the proper
time $\tau$ of an observer on the wall.  They satisfy
\begin{equation}
 f_{\pm}(r_{\pm})\dot{t}_{\pm}^2(\tau)-\frac{\dot{r}_{\pm}^2(\tau)}{f_{\pm}(r_{\pm})}=1,
\end{equation}
where a dot denotes derivative with respect to $\tau$. 
We take the radial coordinates so that the radius of the bubble is
given by $R=r_+=r_-$ in both outer and inner coordinates.
The evolution of the bubble wall is described by the following equation
\cite{Gregory:2014JHEP,Phys.Rev.D35.1747,Aguirre:2005nt}
based on
 Israel's junction 
condition \cite{NuovoCimento.44}
\begin{eqnarray}
\beta_- - \beta_+ = 4 \pi G \sigma R \equiv \Sigma R,
\label{israel}
\end{eqnarray}
where $\beta_{+} \equiv f_{+} \dot{t}_{+} = \pm \sqrt{ f_+ + \dot{R}^2 }$
and  $\beta_{-} \equiv f_{-} \dot{t}_{-} = \pm \sqrt{ f_- + \dot{R}^2 }$. 
From (\ref{israel}) we find the wall radius satisfies the following
equation similar to an energy conservation equation of a particle in a
potential $V(z)$.
\begin{equation}
\left( \frac{dz}{d \tau'} \right)^2\!\!\! + V(z) = E,~V(z) \equiv - \frac{1}{1-s} \frac{\gamma^2}{z} -\left( \frac{1-z^3}{z^2} \right)^2\!\!,
 \label{israel2}
\end{equation}
\begin{equation}
E \equiv -\frac{\gamma^2}{\left[2GM_+ \chi(1-s)\right]^{\frac{2}{3}}},~
\chi \equiv (H^2 + \Sigma^2)^{\frac{1}{2}}, \gamma \equiv \frac{2 \Sigma}{\chi}.
 \label{E12}
\end{equation}
Here dimensionless coordinate variables are defined by 
\begin{eqnarray}
\tau' \equiv \frac{\chi^2 \tau}{2 \Sigma},~ z^3 \equiv \frac{\chi^2 R^3}{2GM_+
(1-s)},~{\rm with}~ s \equiv \frac{M_-}{M_+}.
\end{eqnarray}

\begin{figure}[t]
  \begin{center}
    \includegraphics[keepaspectratio=true,height=140mm]{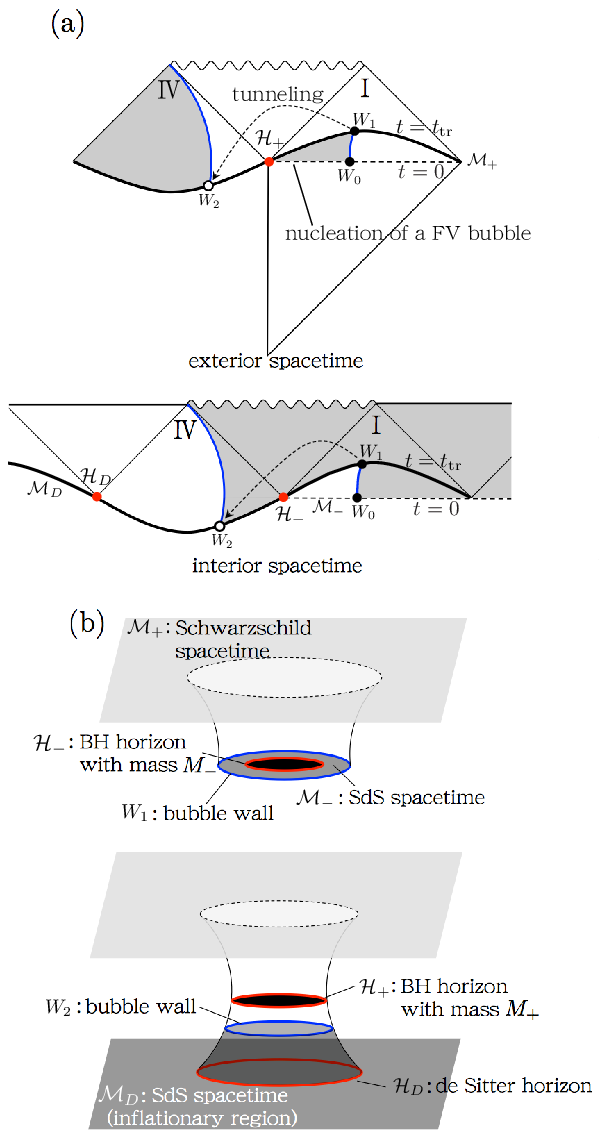}
  \end{center}
  \caption{
The trajectories of a bubble wall (blue line) on Penrose diagrams and a
 schematic figure of a wormhole-like configuration 
accommodating an inflationary region induced by a phase transition.
The upper (lower) diagram in Figure (a) shows the spacetime outside (inside)
the wall and a shaded region is to be replaced by the interior
 (exterior) spacetime.
These diagrams depict the case a bubble wall is produced at the point
 $W_0$ at $t=t_w<0$ through thermal effects of Hawking radiation, and the
 wall tunnels from $W_1$ to $W_2$ at $t=0$ to create a
 wormhole-like configuration.
Figures (b) depict the initial and final configurations schematically.}%
  \label{fig3.5}
\end{figure}
As is seen in Fig.\ \ref{112103}, the potential $V(z)$ has a concave
shape with the maximum $V(z_m)\equiv V_{\max}$ given by
\begin{equation}
V_{\max} = -3\frac{z_m^6-1}{z_m^4}, \label{VTOP12} 
\end{equation}
\begin{align}
&\text{with} \nonumber \\
&z_m^3 = \left[ 2 +\! \left( \frac{1}{2}-\frac{\gamma^2}{4(1-s)} \right)^2 \right]^{\frac{1}{2}}\!\!\!\! - \left( \frac{1}{2}-\frac{\gamma^2}{4(1-s)} \right),  \label{zm}
\end{align}
for $s<1$.  
From (\ref{israel}) we also find
\begin{equation}
M_+=M_-+\frac{4\pi}{3}R^3\epsilon^4+4 \pi R^2 \sigma \frac{\beta_+ + \beta_-}{2} . \label{10}
\end{equation}
We may consider the evolution of the system taking the initial condition
that the bubble is at rest at $R=r_w$ as the Hawking temperature has
increased to above $m$ so that thermal support on the wall has become
less important as discussed above.  
For a thin wall bubble, (\ref{10}) reads $M_+ \cong M_- +4\pi
r_w^3\epsilon^4/3$, and one can show that an inequality 
$E=V(z_w) < -\gamma^2/(1-s)=V(z=1)$ holds where 
$z_w \equiv \chi^{2/3}r_w[H^2 r_w^3+\Sigma r_w^2 (\beta_+ +
\beta_-)]^{-1/3}$
 is the value of $z$ corresponding to $R=r_w$.  Therefore 
 $\beta_\pm$ are both positive
initially as is shown in Fig.\ \ref{112103}.  

We can discuss quantum tunneling of the bubble wall from $R=r_w$
 to a larger $R$, from which it expands in real time,
 by manipulating the Euclidean action.  As one can see in Fig.\ \ref{112103},  
physically relevant expanding bubble nucleation is possible only 
for $\beta_+ <0$ and $\beta_-<0$.
It has been shown in \cite{Phys.Rev.D35.1747} that in this
case the trajectory of the bubble wall after the transition exists in region IV
on the Penrose diagram (Fig. \ref{fig3.5}-(a)), that is, a wormhole-like
configuration is created  \cite{Berezin:1987ea,Ansoldi:2014hta}  and the false vacuum
exists on the other side of the throat (Fig. \ref{fig3.5}-(b)).

Note that although $\beta_+$ and $\beta_-$ change their signs at
different radii, namely $z=1$ and $z_c$, their physical separation
\beq
  \Delta R_{c1}=\left[\frac{2GM_+(1-s)}{\chi^2}\right]^{1/3}\left(
z_c^{1/3}-1\right)\cong\frac{\gamma^2}{4}r_w
\eeq
is actually smaller than the width of the wall $1/m$ for realistic
values of parameters, so that we can regard that they change sign at the
same radius in the thin wall approximation.

Let us now calculate the transition rate to the wormhole-like configuration
$\Gamma$ by solving Euclidean equation of motion starting
from the bubble radius $R=r_w$ at rest.
 Following  Gregory, Moss, and Withers
\cite{Gregory:2014JHEP,Phys.Rev.Lett.115.071303,JHEP.1508.114},
the transition rate $\Gamma$ has the form
\begin{equation}
\Gamma =  m e^{-I_{\calM-\calB}-I_{\calB}}=
  m e^{-B_{\text{tunnel}} + \Delta S},
\label{082001}
\end{equation}
where the prefactor $m$ has been introduced on dimensional grounds.
Here $I_{\calM-\calB}$ represents the action over the regular bulk Euclidean
spacetime and $I_{\calB}$ stands for the contribution of conical
singularities. 
They are given by
\if
 $B_{\text{tunnel}}$ can be calculated by following
the Farhi-Guth-Guven/Fischler-Morgan-Polchinski tunneling \cite{Farhi:1989yr,Fischler:1989se,Fischler:1990pk},
and $\Delta S$ indicates the difference between the Bekenstein entropies before and after the transition.
The two factors $B_{\text{tunnel}}$ and $\Delta S$ have the forms
\fi
\begin{equation}
\begin{split}
I_{\calM-\calB} &= \int d\tau_E  \left[
(2R-6GM_+)\dot{t}_{E+}\right.  \label{Icirc} \\
&\left.~~~~~ -(2R-6GM_-)\dot{t}_{E-}
\right] \equiv B_{\text{tunnel}},\\
I_{\calB} &= \frac{A_f}{4 G} - \frac{A_i}{4G} \equiv \Delta S,
\end{split}
\end{equation}
respectively,
where the suffix $E$ indicates the Euclidean time and
$A_i$ ($A_f$) denotes
the total horizon area in the initial (final) state.
Obviously terms arising from conical singularities are identical 
to the difference of horizon (Bekenstein) entropies, $\Delta S$,
between initial and final states.  These terms have been derived using 
another method of calculation, too 
\cite{Farhi:1989yr,Fischler:1989se,Fischler:1990pk,Chen:2015ibc}.

It is well known that the Bekenstein entropy of horizon may be related to
its number of microscopic states $W$ although so far we do not know
what the microscopic degrees of freedom are.
In our case, the initial state before tunneling has a
Schwarzschild de Sitter black hole horizon with its mass parameter
$M_-$, whose area is denoted by $A_-$,
and the final state has two gravitational horizons, namely,
 the black hole horizon with mass
$M_+$ and the de Sitter horizon (Fig. \ref{fig3.5}-(b)), whose horizon areas are
denoted by $A_+$ and $A_D$, respectively.
Therefore, we have $A_i = A_-$ and $A_f = A_+ + A_D$ and the numbers of
the initial and final microscopic degrees of freedom are given by $W_i = e^{A_i/4G}$
and $W_f = e^{A_f/4G}$, respectively \cite{Chen:2017suz}.

From $e^{\Delta S}=W_f/W_i$ we can interpret 
the transition rate we have calculated, (\ref{082001}), as
a transition from one microscopic initial state with a statistical
weight $1/W_i$ to a final state with $W_f$ microscopic degrees of
freedom, and the transition rate from one microscopic state 
of the initial black hole
to another microscopic state of the final wormhole configuration
is given by
\begin{equation}
\Gamma_{\text{micro}} = m e^{-B_{\text{tunnel}}},
\label{081601}
\end{equation}
up to the uncertainty of the prefactor.

Let us evaluate the transition rate by calculating $B_{\text{tunnel}}$
and $\Delta S$  
which are functions of  $q$,
the energy scales $m$ and  $\epsilon$, and the tension of bubble wall $\sigma$.
Here we can evaluate the tension as $\sigma \simeq \xi^4 / m$,
where $\xi^4$ is the potential energy density at the top of the 
potential barrier separating
the false vacuum and true vacuum. 
We take $M_+$ at a reference value $M_+=\mpl^2/(8\pi m) \gg M_{\text{Pl}}$
corresponding to  $T_H=m$. 
Taking $m^2=120\sqrt{2}\epsilon^2$, $q=1$, and $\xi^4/ \epsilon^4 =25$,
as an example, one can satisfy the thin wall
condition, $mr_w= 120 \gg 1/m$.  $m\Delta R=m\gamma^2 r_w/4 \ll
1$ is also satisfied for $\epsilon \ll 10^{16}$ GeV.  

\begin{figure}[h]
  \begin{center}
    \includegraphics[keepaspectratio=true,height=56mm]{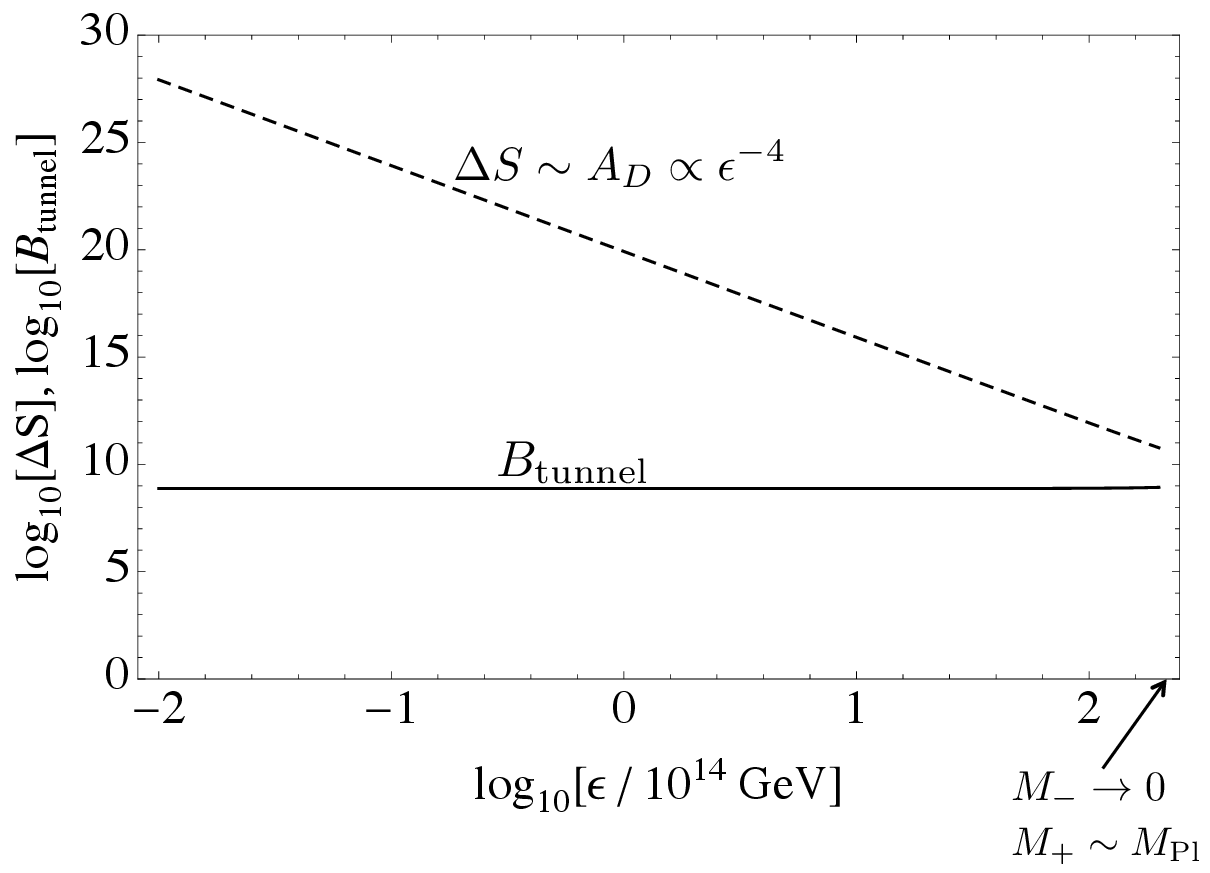}
  \end{center}
  \caption{$\Delta S$ (dashed line) and
  $B_{\text{tunnel}}$ (solid lines) as  functions of $\epsilon$
for $m^2=120\sqrt{2}\epsilon^2$, $q=1$ and $\xi^4/\epsilon^4 = 25$.
The inner black hole mass, $M_-$, approaches to zero and $M_+$ becomes comparable to the Planck mass
at $\epsilon \simeq 2\times 10^{16}$ GeV, where $M_{+} = 4\times 10^2\mpl (\epsilon/10^{14}{\rm GeV})^{-1}$.
  }%
  \label{fig3}
\end{figure}

Figure 3 depicts $\Delta S$ and $B_{\text{tunnel}}$
as  functions of $\epsilon$. 
$\Delta S$ is proportional to $\epsilon^{-4}$ since the de Sitter horizon area, which is proportional to $H^{-2} \propto \epsilon^{-4}$, becomes dominant compared to the black hole horizon areas for $\epsilon \ll 10^{16}$ GeV. 
As is seen here, we always find $\Delta S \gg B_{\text{tunnel}} \gg 1$.
This means that even though the tunneling rate from one microscopic
state to another is exponentially suppressed so that the semiclassical
approximation is valid, due to the largeness of the number of 
microscopic degrees of freedom after the transition, the tunneling 
as a whole is unsuppressed and wormhole creation may take place with the relevant time scale
$t \sim 1/T_H \sim 1/m$, once a bubble is thermally excited around an evaporating black
hole with the proper conditions we discussed above.  Similar enhancement
of transition rate due to the large entropy in the final state has been
observed by Mathur in a different problem \cite{Mathur:2008kg}.

Then we can sketch the following scenario of cosmic evolution.
Typical astrophysical black holes with mass $\sim 10M_{\odot}$ will
evaporate in $\sim 10^{67}$ years from now.  As its mass falls below a
critical value so that the Hawking temperature become high enough for
 a false vacuum bubble to spontaneously nucleate around the black hole
according to the process described by Moss \cite{Moss:1984zf}.
Then the  bubble wall will experience quantum tunneling rather efficiently to 
create a wormhole-like configuration with a de Sitter horizon, beyond
which the false vacuum region is extended to infinity.
Thus the space on the other
side of the throat will inflate which is
causally disconnected from our patch of the universe.  If inflation is
appropriately terminated followed by reheating, another big bang
universe will result there.  For this purpose the old inflation model 
\cite{Sato:1980yn,Guth:1980zm} with thin wall bubble nucleation does not
work, but we may make use of the results of open inflation models there
\cite{Sasaki:1993ha,Yamamoto:1995sw,Bucher:1994gb}
which can also realize an effectively flat universe.  

Throughout these processes, the outer geometry remain Schwarzschild space with the mass
parameter $M_+$, so those who live there do not realize a black hole in
 their universe
has created a child universe.
To this end alone, our model is similar to the scenario
proposed by Frolov, Markov, and Mukhanov  \cite{Frolov:1988vj,Frolov:1989pf}.
However, there are two striking differences between our model
and their scenario.  One is that theirs is entirely
dependent on the limiting curvature hypothesis and the assumption
that in the regime of large curvature, the gravitational field equation 
would take the form in vacuum with a positive cosmological 
constant. They thereby find a Schwarzschild solution is continued
to a deflating de Sitter space inside the black hole horizon
which bounces to an inflating de Sitter universe.  Our model, on the
other hand, does not need such a speculative hypothesis near the 
singularity but creation of another inflationary universe is achieved
by symmetry restoration due to the high Hawking temperature
around an evaporating black hole which also induces a phase transition 
to produce a wormhole-like configuration in quantum field theory.
Thus the entire processes can be described
by known physics with  appropriate values of the model parameters.
Another difference lies in the causal structures as described
in Fig. \ref{fig3.5} of our paper and Figs. 3 and 6 of \cite{Frolov:1988vj},
that is, in our model the inflating domain is causally disconnected from the original universe
unlike theirs.

In conclusion, our result may also suggest that our Universe may have been
created from a black hole in the previous generation in the cosmos.

\textit{Acknowledgements.}
We would like to thank Daisuke Yamauchi, Teruaki Suyama, and Dong-han Yeom for helpful discussions. 
This work was partially supported by JSPS Grant-in-Aid for Scientific Research
15H02082 (J.Y.), Grant-in-Aid for Scientific Research on Innovative Areas No.
15H05888 (J.Y.), Grant-in-Aid for JSPS Fellow No. 16J01780 (N.O.),
and a research program of the Advanced Leading Graduate Course for
Photon Science (ALPS) at the University of Tokyo (N.O.).

\end{document}